# Pinning a Domain Wall in (Ga,Mn)As with Focused Ion Beam Lithography

A. W. Holleitner, H. Knotz, R. C. Myers, A. C. Gossard and D. D. Awschalom*

*Center for Spintronics and Quantum Computation, University of California, Santa Barbara, California 93106, USA*

* Correspondence should be addressed to D.D.A. E-mail: awsch@physics.ucsb.edu

We utilize a focused beam of $Ga^+$ ions to define magnetization pinning sites in a ferromagnetic epilayer of (Ga,Mn)As. The nonmagnetic defects locally increase the magneto-crystalline anisotropy energies, by which a domain wall is pinned at a given position. We demonstrate techniques for manipulating domain walls at these pinning sites as probed with the giant planar Hall-effect (GPHE). By varying the magnetic field angle relative to the crystal axes, an upper limit is placed on the local effective anisotropy energy.

Ferromagnetic semiconductors such as p-type $(Ga,Mn)As$ are well suited for investigating the interplay between magnetic domains and electronic carriers, since both the growth of III-V semiconductor heterostructures and the fabrication of electrical devices are well established via molecular beam epitaxy (MBE) and nano-lithography, respectively[1,2]. In order to pin and probe magnetization domains at a given position on a Hall-bar, the effective anisotropy energies of the crystal need to be controlled on a mesoscopic level. Studies on Co, Pt, and Fe [3-6] demonstrated that variations of the film thickness even down to the monolayer scale can pin a magnetization domain wall (DW) due to local reduction of the demagnetization field. Here we use a focused ion beam (FIB) of $Ga^+$ ions to imprint domain pinning centers at specific locations along a patterned Hall mesa of a $(Ga,Mn)As$ epilayer. We show that these nonmagnetic defects increase the local magnetic anisotropy energy, while the transport properties of the ferromagnetic epilayer are qualitatively unchanged. This flexible approach to control domains provides a pathway for sensitive DW experiments[7], magnetic logic-circuits[8] and future experiments on the macroscopic tunneling of DWs[9].

Epilayers of $Ga_{0.947}Mn_{0.053}As$ with a thickness $t=100\ nm$ and a hole density of about $6\times10^{19}\ cm^{-3}$ ($4.2\ K$) are grown by MBE at $250\ °C$ in a Varian GenII system optimized for low temperature growth[10]. The magnetic layer is deposited on $400\ nm$ of $Al_{0.4}Ga_{0.6}As$ grown at $585\ °C$ on a semi-insulating $GaAs$ (001) substrate. The magnetic properties are first characterized by a SQUID magnetometer. The in-plane easy axes are found to be close to [100], and the saturation magnetization and the Curie temperature are $M_S \sim (2.1\pm 0.1)\ \mu_B/Mn^{2+}$ and $T_C \sim 71\ K$, respectively. Using standard wet-etching techniques, Hall bars are photo-lithographically defined with a width of $w=100\ \mu m$, an aspect ratio of $4:1$, and Hall probe sizes ranging between $(1-100)\ \mu m$. The Hall bars are patterned parallel to one of the magnetic easy axes. The white arrow in the micrograph of Fig. 1(a) highlights a $Ga$-pinning line formed by the FIB which lies in-plane along [110]. The lines, written with an acceleration voltage of $U = 30\ kV$, a current $I_{Ga^+} = 1\ pA$, and a lithographic width of $(60-120)\ nm$, have no qualitative effect on the longitudinal resistance of the epilayers, which is consistent with the nonmagnetic pinning sites only being incorporated close to the surface of the epilayer. A standard lock-in technique with an AC-bias of $I_{RMS} = (5-500)\ nA$ at $f = 18\ Hz$ along the Hall bar directions enables the measurement of the resistances $R_A$, $R_B$, and $R_C$ from $T = 260\ mK$ up to room temperature (Fig. 1(b)).

Highlighting the fact that the hole-mediated transport properties of $(Ga,Mn)As$ are strongly correlated to its ferromagnetic characteristics, Fig. 2(a) shows longitudinal resistance as a function of magnetic field for sample 1. The data exhibits sharp switching features due to the longitudinal GPHE.[11] At high positive field the magnetization saturates along $\vec{M}_1 \parallel [100]$ (green) which is parallel to the axis of the Hall bar. As the field switches sign, the magnetization first becomes metastable and then realigns along the perpendicular cubic $[0\bar{1}0]$ axis (yellow) which in turn causes an abrupt change of the longitudinal resistance (red curve). Further increasing the applied negative field leads to a saturation magnetization along $\vec{M}_3 \parallel [\bar{1}00]$ (blue). Again, the transport data reveal an abrupt jump back to a resistance value comparable to that observed for the previous saturation field. This symmetry also appears if the field is swept in the opposite direction (black curve), which leads to an identical resistance jump at the field strength of opposite sign. Generally, the change in resistance observed for $\vec{M}_2$ is described by the Hall-field given by the anisotropic magnetoresistance,

$$\vec{E} = (\rho_\parallel - \rho_\perp)(\hat{m} \cdot \vec{j})\hat{m}, \tag{1}$$

where $\hat{m}$ is the unit vector of the in-plane magnetization, $\vec{j}$ is the current vector, and $(\rho_\parallel - \rho_\perp)$ is the difference between the anisotropic magnetoresistivities parallel and perpendicular to the current direction[12]. The inset of Fig. 2(a) illustrates the temperature dependence of the magnitude of the resistance jump. Typical for the GPHE in $(Ga,Mn)As$, the jump magnitude shows a rapid increase as the temperature is lowered. The saturation value at low temperature of about $100\,\Omega$ translates into $(\rho_\parallel - \rho_\perp) = -2.5 \times 10^{-6}\,\Omega m$. For comparison, we typically find a bulk resistivity of $\rho \sim 8.4 \times 10^{-5}\,\Omega m$ (at $4.3\,K$).

In $(Ga,Mn)As$ epilayers, changes in the global magnetization are driven by the passing of a $90°$-DW[11,13]. Fig. 2(b) demonstrates how a DW is caught at the pinning line. After saturation at large positive fields, the magnetic state $\vec{M}_1$ is metastable at small negative fields. At $\vec{H} = -8.6\,mT$, $R_C$ jumps into the more resistive state $\vec{M}_2$, followed by $R_A$ at $\vec{H} = -9.2\,mT$. After $R_C$ switches, $R_A$ features a small resistance increase of $\sim 0.05\,\%$ (red circle) which is due to a local magnetization reversal at the pinning line. Since the occurrence of this type of resistance increase is found to be independent of the Hall probe size, and is only observed with Hall probes traversed by a pinning line, we conclude that a DW is pinned as depicted in Fig. 2c (the arrows indicate the $Mn^{2+}$-spin orientation for $\vec{M}_1$ and $\vec{M}_2$) and that the extra resistance is not due to geometrical constrictions[14,15]. In principle, the resistance increase also occurs for jumps from $\vec{M}_2$ to $\vec{M}_3$. However, the jump in $R_A$ for $\vec{M}_2$ to $\vec{M}_3$ exhibits a long saturation tail towards a higher negative field. This reflects the fact that the effective anisotropy energies at the pinning line are enhanced compared to the bulk anisotropy energies.

Subtracting the background resistance value $R_B$ from $R_A$ yields the influence of the pinning line with greater resolution. Fig. 3(a) depicts both resistances $R_A$ and $R_B$, clearly demonstrating that $R_A$ is offset by a constant resistance of $75.3 \pm 0.3\,\Omega$. Depending on the particular device, this offset varied in the range of 10 to $100\,\Omega$ and is dominated by the resistive influence of the pinning defects generated by the FIB. In Fig. 3(b), the resistance offset defines the amplified resistance level of $R_A - R_B$ for the saturated magnetization $\vec{M}_1$ (green). As stated above, at $\vec{H} = -8.6\,mT$ a DW becomes pinned. The signal then exhibits an extra resistance increase $\Delta R$ (encircled data) which includes the resistance of a DW[7]. Subsequently, just before the magnetization jumps from $\vec{M}_2$ (yellow) to $\vec{M}_3$ (blue) a resistance increase again occurs (up arrow).

Generally, the orientation of the magnetization in $(Ga,Mn)As$ epilayers can be described by cubic and uniaxial anisotropy energy densities $K_C$ and $K_U$, respectively. In accordance with ref. [11], we find that the uniaxial symmetry axis is parallel to a hard cubic axis. By applying the magnetic field $\vec{H}_{ext}$ at an angle $\theta_H$, the magnetization $\vec{M}$ orients along the angle $\theta$ such that the following free energy is minimized,

$$E = K_U \sin^2(\theta - \pi/4) + \frac{K_C}{4}\sin^2(2\theta) - \vec{M}\cdot\vec{H}_{ext}. \qquad (2)$$

At low magnetic field amplitudes, the magnetization is oriented along the minima of this expression which are parallel to <100> with a deviation of $\Delta_{MIN} = \pm 1/2 \arcsin(K_U/K_C)$. Fig 3(c) depicts the longitudinal resistance as the in-plane magnetic field angle is swept at constant field magnitudes. Whenever the external magnetic field forces the magnetization to pass a hard axis of the crystal, the resistance exhibits a hysteretic switching depending on the field being swept clockwise or counter clockwise. By fitting the data to equation (2) we extract the anisotropy fields of the crystal to be $H_C = 2K_C/M = 2.5 \times 10^5\, Am^{-1}$ and $H_U = 2K_U/M = 8.8 \times 10^3\, Am^{-1}$. The hysteresis can be detected for external magnetic flux densities as high as the dominating cubic anisotropy field, which corresponds to $H \cong 300\, mT$ and is indicated by the red bold curve. For the highest field magnitudes, however, the magnetization smoothly follows the direction of the external magnetic field. In this regime the longitudinal resistance change is described by equation (1) with a uniformly rotating unit vector $\hat{m}$.

Figure 3(d) displays the field angle dependence data for the resistance difference $R_A$-$R_B$. Since $R_A$ spans the FIB-line, this measurement is sensitive to the local misalignment of the Mn spins in the pinning line. For low magnetic fields, we see switching events which resemble the ones of Fig 3(b) and (c) (for clarity we only show data for sweeping the magnetic field counter clockwise). However, the asymmetry in the curve (dashed line) persists even to field ranges as high as $H = 450\, mT$, which is 50% above the dominating cubic anisotropy field measured in Fig. 3(c). Even at this magnitude, some of the Mn spins in the pinning line are still misaligned with respect to the external field.

We deduce that the effective anisotropy fields within the pinning line are locally increased by up to 50%.

In conclusion, a FIB technique is utilized in order to manipulate and to probe magnetic domains in homogenous epilayers of $(Ga,Mn)As$. The FIB induces nonmagnetic defects which locally increase the effective anisotropy energies by up to 50 %. By using the anisotropic magnetoresistance as a read out of the magnetization states, we demonstrate how a domain wall can be accurately centered in a multi-probe transport circuit to allow high precision domain wall experiments.

We thank R. J. Epstein and F. Meier for enlightening discussions and J. Davis for technical support. We acknowledge financial support by AFOSR and ONR.

Figure Caption:

FIG. 1(a) Scanning electron micrograph of an epitaxial layer of $(Ga,Mn)As$ featuring a pinning line for macroscopic domain walls (white arrow). (b) Hall circuit with a current $I_{SD}$ along the [100] crystal axis and an in-plane magnetic field $\vec{H}$ at an angle $\theta_H$. Comparison of longitudinal resistances $R_A$ and $R_B$ yields the influence of the pinning line (dashed line).

FIG. 2(a) Longitudinal resistance hysteresis of a device without a pinning line at $\theta_H = (17 \pm 1)°$ and $4.3\,K$ reflects the magnetization states $\vec{M}_1$ (green), $\vec{M}_2$ (yellow) and $\vec{M}_3$ (blue) along the cubic in-plane axes for the red curve. Inset: temperature dependence of the resistance jump at low fields. (b) Resistances $R_A$ and $R_C$ at $T = 270\,mK$ demonstrating the pinning of a DW (red circle) in sample 2. Resistance $R_A$ is higher due to the resistive influence of the pinning line (for clarity $R_C$ is artificially offset by $-20\,\Omega$). c, Magnetization sketch for a $90°$-DW between $\vec{M}_1$ and $\vec{M}_2$, centered at the pinning line (red dashed line).

FIG. 3(a) Measuring $R_A$ vs magnetic field senses the influence of the pinning line (red circle), whereas $R_B$ only monitors the resistance of the bulk epilayer. (b) Amplifying $R_A - R_B$ shows the resistance increase due to a DW (red circle) between the magnetization states $\vec{M}_1$ (green) and $\vec{M}_2$ (yellow). The effective anisotropy energies are extracted by the angular field dependence of the longitudinal resistance (c) of sample 3 without any pinning line and in sample 4 of the resistance difference $R_A - R_B$ sensing a pinning line (d). Curves are displayed with an offset for clarity (field magnitudes from bottom to top: 50 mT, 100 mT, …, 500 mT).

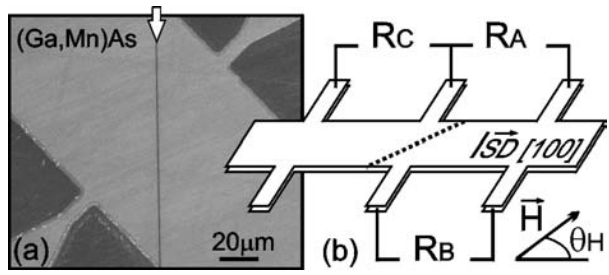

Holleitner et al. FIG. 1

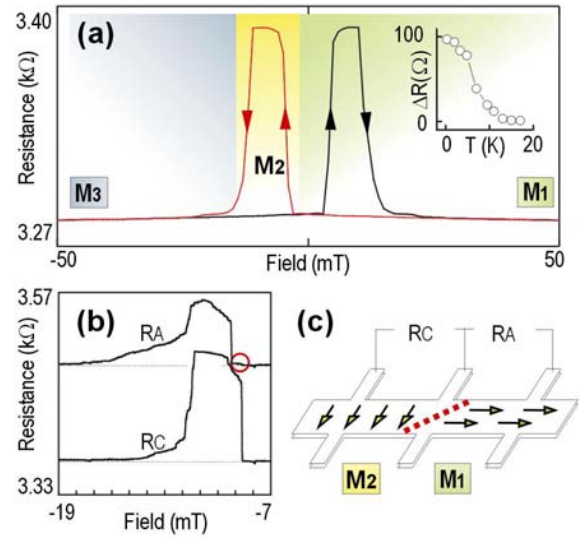

Holleitner et al. FIG. 2

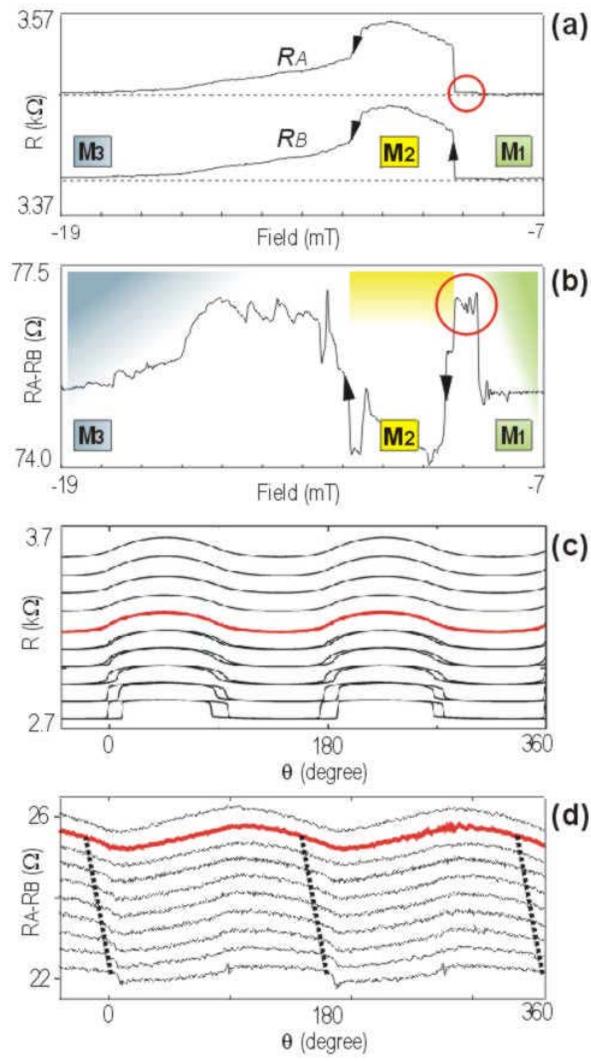

Holleitner et al. FIG. 3